\documentclass[aps,prl,twocolumn,showpacs,superscriptaddress,groupedaddress]{revtex4}  
\usepackage{graphicx}  
\usepackage{dcolumn}   
\usepackage{bm}        
\usepackage{amssymb}   
\usepackage{hyperref}
\usepackage{color}

\begin{document}

\title{Identification of intermittent multi-fractal turbulence in fully kinetic simulations of magnetic reconnection}

\author{E. Leonardis}
\email{e.leonardis@warwick.ac.uk}
\affiliation{Centre for Fusion, Space and Astrophysics, Department of Physics, University of Warwick, Coventry, CV4 7AL, United Kingdom}
\author{S. C. Chapman}
\affiliation{Centre for Fusion, Space and Astrophysics, Department of Physics, University of Warwick, Coventry, CV4 7AL, United Kingdom}
\author{W. Daughton}
\affiliation{Los Alamos National Laboratory, Los Alamos, New Mexico 87545, USA}
\author{V. Roytershteyn} \thanks{Now at Sciberquest, Inc, Del Mar, California	92014, USA}
\affiliation{University of California, San Diego, La Jolla, California 92093, USA}
\author{H. Karimabadi}
\affiliation{University of California, San Diego, La Jolla, California 92093, USA}

\date{\today}

\begin{abstract}
Recent fully nonlinear, kinetic three-dimensional simulations of magnetic reconnection \citep{Daughton2011} evolve structures and exhibit dynamics on multiple scales, in a manner reminiscent of turbulence. These simulations of reconnection are among the first to be performed at sufficient spatio-temporal resolution to allow formal quantitative analysis of statistical scaling which we present here. We find that the magnetic field fluctuations generated by reconnection are anisotropic, have non-trivial spatial correlation and exhibit the hallmarks of finite range \emph{fluid} turbulence; they have non-Gaussian distributions, exhibit Extended Self-Similarity in their scaling and are spatially multifractal. Furthermore, we find that the field $\bf J \cdot E$ is also multifractal, so that magnetic energy is converted to plasma kinetic energy in a manner that is spatially intermittent. This suggests that dissipation in this sense in collisionless reconnection on kinetic scales has an analogue in fluid-like turbulent phenomenology, in that it proceeds via multifractal structures generated by an intermittent cascade.
\end{abstract}

\pacs{94.30.cp 94.05.Lk}

\maketitle
Magnetic reconnection is a fundamental process that converts magnetic energy into various forms of plasma kinetic energy. It is thought to occur in a variety of space, astrophysical and laboratory applications, with parameter regimes spanning from collisional to highly collisionless plasmas (e.g. see Ref.~\cite{ji2011} and references therein). While many studies have focused on laminar initial conditions, it is now widely recognized that the influence of turbulence remains a major uncertainty.   Depending on the application, the turbulence may arise from a spectrum of instabilities within the reconnection layer or from pre-existing magnetic fluctuations in the ambient plasma. Within the magnetohydrodynamic (MHD) model, there has been progress on both ideas - either by starting with a laminar current sheet to explore instabilities \cite{loureiro07,lapenta08,bhattacharjee09,huang10} or by directly driving turbulence \cite{Servidio2011,matthaeus86,Kowal2009,Loureiro2009,eyink11}.     

Moving beyond the MHD model into kinetic regimes, most research has focused on initially laminar current sheets within a variety of descriptions \cite{birn01} including two fluid, hybrid and fully kinetic simulations, which allow a complete description of the electron physics responsible for breaking the frozen-flux condition in collisionless parameter regimes \cite{hesse99,pritchett01}. As larger kinetic simulations became possible, one surprising result was that the nonlinear evolution of reconnection produced extended electron-scale current sheets, with half-thickness on the order of the electron inertial length and lengths that can extend beyond the ion inertial scale \cite{daughton06,drake06b,karimabadi07,klimas08,shay07}. These predictions have since been confirmed in spacecraft observations \cite{Phan2007}. While the precise details depend on the strength of the guide field (i.e. magnetic shear angle), it has been demonstrated that electron pressure anisotropy plays a key role in setting up and driving these layers \cite{ohia12}.  Thus the existence of these structures is now well accepted and a variety of two-dimensional (2D) kinetic simulations have shown the layers can become unstable to secondary magnetic islands \cite{daughton06,drake06b,karimabadi07,klimas08} leading to time dependent scenario.  Recent extensions of these kinetic simulations to 3D have demonstrated that the tearing instability within these electron layers has much greater freedom to develop and gives rise to numerous magnetic flux ropes \cite{Daughton2011}. The subsequent nonlinear interaction of these flux ropes is seen in the simulations to lead to the self-generation of structures on multiple scales within the initially laminar ion-scale current layer. The question then arises - is the multi-scale nature of these flux ropes important for dissipation in the sense of conversion of magnetic to kinetic energy?
The key properties of turbulence in this context are that it cascades energy in a scale invariant and intermittent manner. Thus turbulence provides a mechanism to form a spatial multifractal field of coherent structures across a broad range of spatial scales which can then contribute to dissipation of magnetic field energy into plasma kinetic energy.

In this Letter we show that these fully kinetic 3D PIC simulations of magnetic reconnection \cite{Daughton2011} do indeed exhibit the hallmarks of intermittent turbulence.
A key property of turbulence is that it can be characterized and quantified in a robust and reproducible way in terms of ensemble averaged statistical properties of fluctuations. In an infinite medium, fully developed turbulence exhibits statistical \textit{scale invariance} in fluctuations in the bulk quantities that describe the flow. Either when the turbulence is not completely evolved (low Reynolds number), or the system is of finite size, a generalized scale invariance or extended self similarity (ESS) \cite{Benzi1993} still holds in both hydrodynamic and magnetohydrodynamic (MHD) turbulent flows, as seen for example, in the solar wind \cite[e.g.,][]{Bruno&Carbone,Chapman2009a,Chapman2009b} and in a solar quiescent prominence \cite{Leonardis2012}.
A key discriminator of turbulence is that fluctuations have non Gaussian probability density functions (PDFs) with moments that exhibit multifractal scaling. 
Remarkably, we find that the magnetic field fluctuations on kinetic scales in these 3D simulations of reconnection have this quantitative statistical character, specifically, they show non-Gaussian statistics and ESS consistent with multifractal scaling. Furthermore, dissipation, in the sense of conversion of magnetic to kinetic energy, occurs on a multifractal field. 
This suggests that dissipation in collisionless reconnection on kinetic scales has an analogue in fluid-like turbulent phenomenology, in that it proceeds via multifractal structures generated by an intermittent cascade.

We focus on 3D simulations of undriven magnetic reconnection in collisionless plasmas \cite{Daughton2011}. These petascale simulations use the kinetic particle-in-cell code VPIC, which solves the full set of relativistic Maxwell's equations. The simulation is initialized with a Harris current sheet, with magnetic field $\mathbf{B} = B_{xo} tanh (z/ \lambda) \mathbf{e_x}+ B_{yo} \mathbf{e_y}$, where $\mathbf{e_x}$ and $\mathbf{e_y}$ are unit vectors, $B_{yo} = B_{xo}$ is a uniform guide field and $\lambda = d_i$ is the initial half-thickness, where $d_i$ is an ion inertial length. Here the ion electron mass ratio is $m_i/m_e=100$, which implies that $d_i$ is 10$d_e$, the electron inertial length.
The simulation employs open boundary conditions in the $x$ and $z$ directions and periodic boundary conditions in the $y$ direction. The dimensions of the simulation grid are $L_x \times L_y \times L_ z = 2,048 \times 2,048 \times 1,024$ cells, corresponding to $70 d_i \times 70 d_i \times 35 d_i$. 
We consider a slice of the 3D simulation grid in the X-Z plane at Y = 35$d_i$ and at the time $t \Omega_{ci} = 78$ of the simulation, where $\Omega_{ci} \equiv e B_{x0}/(m_i c)$. This time slice corresponds to a middle phase of the magnetic reconnection in which the turbulence quantitatively reaches its peak of evolution as seen in the power level of the region of scaling in the power spectral density (PSD). The overall time evolution of the simulation is shown in \cite{Daughton2011}. We also analysed an early and late phase of the process at $t \Omega_{ci}$ = 40 and 98 respectively, and find that the reconnection generated fluctuations evolve in time: at early time the power in these fluctuations is weak though above the noise, then it grows in amplitude up to the middle phase where the total integrated field energy density over the turbulent region of the power spectral density is $\sim 0.1$ that of the background field. 
The reconnection rate in these kinetic simulations is fast, $V_{in} / V_A \approx 0.1$, as is the case for kinetic simulations of reconnection where the flow is almost laminar \cite[e.g.][and references therein]{Daughton&Roytershteyn}; hence these kinetic simulations do not show correlation between the level of turbulence fluctuations and the reconnection rate as has been postulated for reconnection on MHD scales \citep{Lazarian1999,Kowal2009}.

\begin{figure}[!ht]
\begin{center}
\includegraphics[scale=0.155]{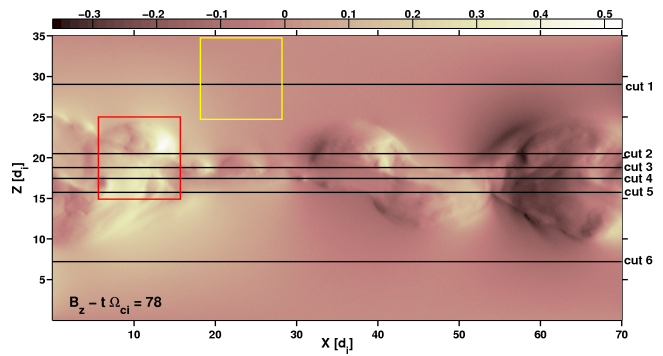}
\caption{$B_z$ in the X-Z plane at Y=35$d_i$ and at the time $t \Omega_{ci} = 78$ of the simulation. Solid black lines are the cuts chosen for the analysis of the magnetic fluctuations, while squares indicate the regions over which we perform the box counting analysis of $\bf J \cdot \bf E$.}
\label{Bzgrid}
\end{center}
\end{figure}
Figure \ref{Bzgrid} shows the $z$-component of the magnetic field, $B_z$, at $t \Omega_{ci} = 78$ corresponding to the time-space slice for which we present analysis here.
We take cuts on the simulation grid along the horizontal direction and label them as cut 1 to 6 (see Fig.~\ref{Bzgrid}).
Cuts 2-5 are within the reconnection region where the field is strongly fluctuating, while cut 1 and 6 lie where there is negligible signature of reconnection. The latter are taken  to quantify the effects of the PIC simulation noise.
Here we focus upon the analysis of the $z$-component of the magnetic field. The reconnection generated fluctuations are highly anisotropic in character and have clearest signature in their $z$-component which is perpendicular to the $x$-$y$ plane of the macroscopic field of these simulations. This anisotropy parallels what has been recently observed both in kinetic range turbulence in the solar wind \cite[e.g.,][]{Kiyani2013,Turner2012} and in a reconnection jet \cite{Huang2012}.

\begin{figure}[!ht]
\begin{center}
\includegraphics[scale=0.155]{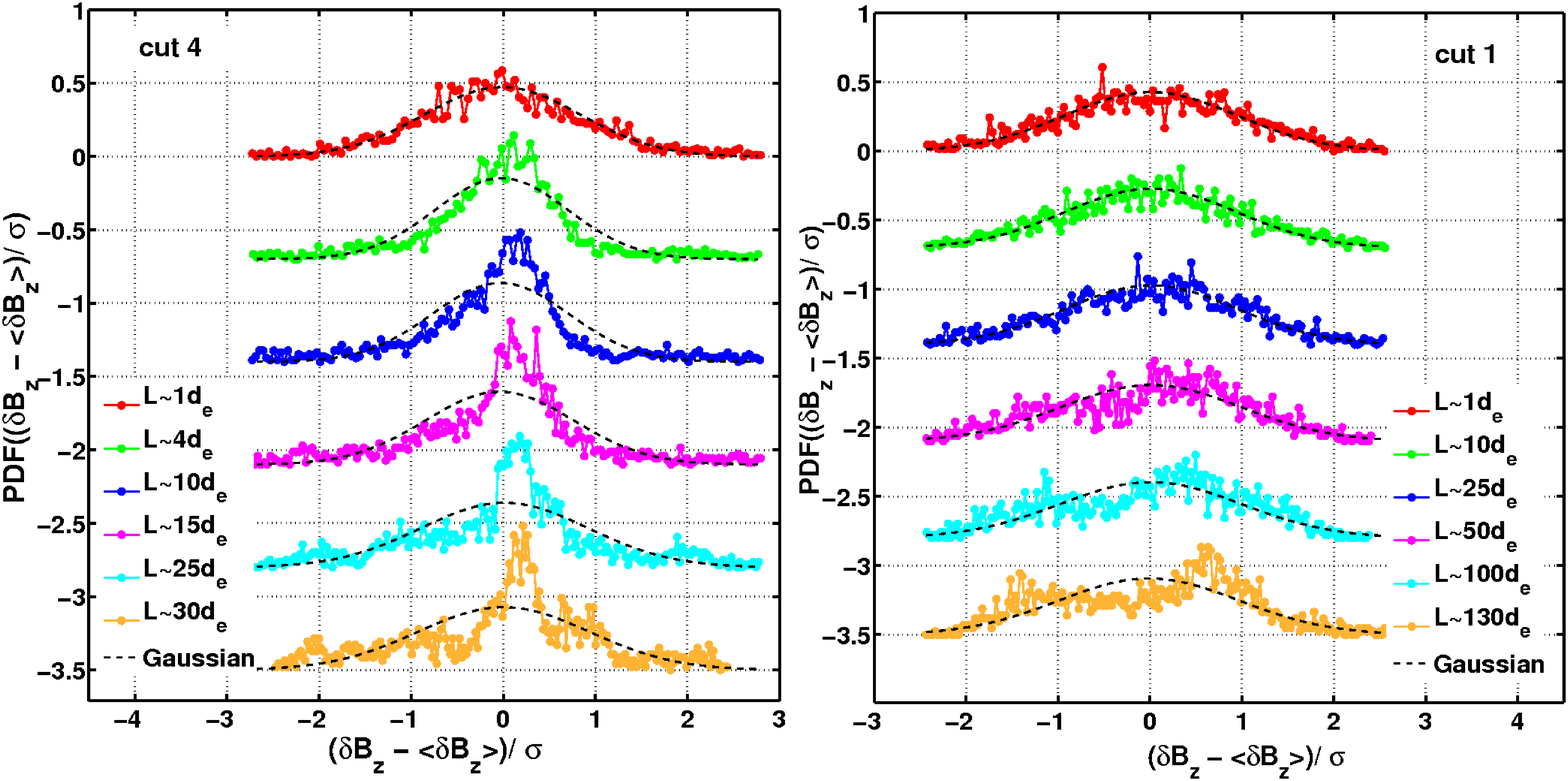}
\caption{PDFs of the fluctuations $\delta B_z(L)$ = $B_z(X+L)-B_z(X)$ of cut 4 (left panel) and cut 1 (right panel) for six different values of the space lag $L$. In both panels, all PDFs are centred on the mean value $<\delta B_z(L)>$ and normalized to the standard deviation $\sigma$ to allow comparisons with a Gaussian distribution (dashed black lines). All curves are shifted along the $y$-direction for clarity.}
\label{PDF}
\end{center}
\end{figure}

Fluctuations associated with fully developed intermittent turbulent flows are characterized by non-Gaussian probability distribution functions (PDFs). Under the assumption of statistical stationarity and homogeneity, fluctuations of a field $I$ on length scale $L$ along a given direction $r$ are defined as $\delta I(L)=I(r + L)-I(r)$.
Figure \ref{PDF} shows the PDFs of the fluctuations $\delta B_z(L)$ at different scales $L$ for cut 4 (left panel) and cut 1 (right panel). We can see that the fluctuations of cut 4 (reconnection generated fluctuations) follow a non-Gaussian distribution at scales $L$ in the range $4d_e<L<25d_e$, while fluctuations associated with cut 1 (noise) are Gaussian distributed up to scales $L \sim 100d_e=10d_i$. The latter suggest that the PIC noise behaves as Brownian noise.
We recover consistent statistics also for the other cuts considered, that is, Gaussian distributions for cut 6 at scales $L$ up to $\sim 10d_i$ and non-Gaussian PDFs for cuts 2,3 and 5 within the range $4 d_e < L < 25d_e$. We identify the latter as a potential range of turbulence, which we now test.

A central characteristic of turbulence is that the PDFs of non-Gaussian fluctuations at different scales are related by a multifractal similarity or scaling. We now test for this by examining the generalized structure functions (GSFs) of the magnetic field fluctuations $\delta B_z(L)$, defined as $ S_p(L) \equiv  < \vert \delta B_z(L) \vert ^p > $, where the angular brackets indicate an ensemble average over $r$, implying approximate statistical homogeneity. In infinite range, fully developed turbulence, one would expect the GSFs to scale as $ S_p(L) \sim L^{\zeta(p)}$, where the $\zeta(p)$ are the scaling exponents.
For  turbulence in a finite domain, or for turbulence that is not fully developed, a generalized scale invariance or ESS can hold, with $S_p(L) \sim G(L)^{\zeta(p)}$, where the function $G(L)$ is an initially unknown function that depends on the largest scale physical structures \cite{Grossmann1994,Bershadskii2007,Chapman2009b}.\\
While for fractal fields $\zeta(p)$ is linear in $p$, intermittent turbulence is realized by multifractal topology (e.g. \cite{Frisch}) and the $\zeta(p)$ are non linear in $p$.
In finite range turbulence, one does not have direct access to the scaling exponents $\zeta(p)$, however it possible to obtain their ratios, $\zeta(p)/\zeta(q)$, by plotting one structure function of order $p$ against another structure function of order $q$. Thus, ESS establishes the following scaling for the structure functions $S_p(L)=\left[S_q(L)\right]^{\zeta(p)/\zeta(q)}$ \cite{Benzi1993}.

\begin{figure}[!ht]
\begin{center}
\includegraphics[scale=0.17]{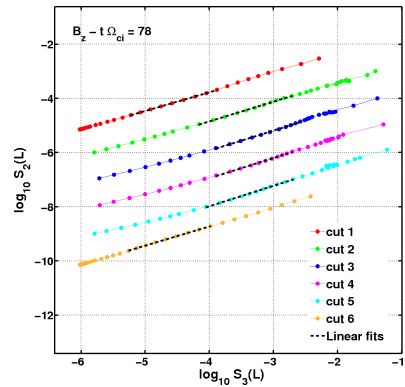}
\caption{Log-log plots of $S_2$ against $S_3$ for all the cuts of interest. Dashed black lines correspond to the linear regression fits in the potentially turbulence range $4 d_e < L < 25d_e$ for cuts 2 to 5 and within the range $1 d_i < L < 10d_i$ for cuts 1 and 6. All curves are shifted along the $y$-direction for clarity.}
\label{ESS}
\end{center}
\end{figure}

\begin{figure}[!ht]
\begin{center}
\includegraphics[scale=0.20]{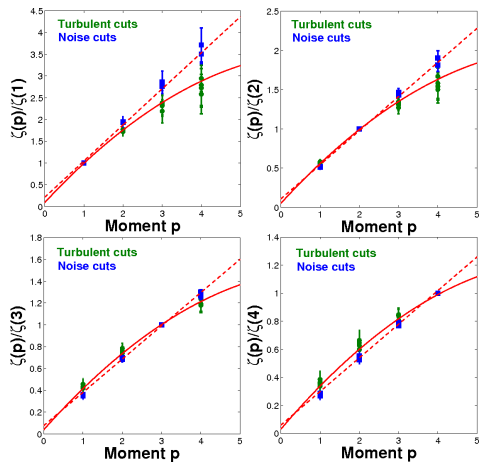}
\end{center}
\caption{$\zeta(p)/\zeta(q)$ versus $p$ for cuts 1 and 6 (blue rectangles) and cuts 2 to 5 (green dots) and for $q$ = 1 (top-left), 2 (top-right), 3 (bottom-left) and 4 (bottom-right). Solid and dashed red lines are the best fits to the curves within uncertainty for the turbulent (cuts 2-5) and noise (cuts 1,6) cuts respectively.}
\label{scalingexponents}
\end{figure}

In Figure \ref{ESS} we plot $S_2$ versus $S_3$ on a logarithmic scale for all the cuts considered. We see that ESS holds for both cuts 1 and 6, which are simulation noise, and cuts 2 to 5, which are within the potentially turbulent range $4 d_e < L < 25d_e$. This implies that both noise and reconnection generated fluctuations have a range of scale invariance.
Now, the question that immediately arises as to whether these fluctuations are multifractal or not. We calculate all possible combinations of the scaling exponents ratio $\zeta(p)/\zeta(q)$ for $p,q = 1, 2,3$ and $4$ by plotting log($S_p$) versus log($S_q$) and by reading the gradients of the linear fits to these curves within the potentially turbulent range, $4 d_e < L < 25d_e$, for cuts 2 to 5 and in the noise range, $1 d_i < L < 10d_i$, for cuts 1 and 6.
The panels in Figure \ref{scalingexponents} show the ratios $\zeta(p)/\zeta(q)$ versus $p$ for $q = 1$ up to $4$ for cuts 1 and 6 (noise) and cuts 2 to 5 (reconnection-fluctuations). The noise cuts (blue rectangles) show a linear behaviour of $\zeta(p)$ with $p$, consistent with a fractal field. The PIC noise thus generates a spatial field of magnetic fluctuations which is a self-affine Brownian noise, which shows fractal scaling. Importantly, it is clearly distinguishable from the reconnection generated fluctuations of cuts 2 to 5 (green dots), which instead consistently show a non linear dependence of $\zeta(p)$ on $p$ within uncertainty. The reconnection generated structures are thus robustly characterized by a multifractal spatial field. At minimum, this suggests a new signature of reconnection outflow regions that could provide a method for observational identification, as has indeed been recently observed \cite{Huang2012}. 
However, this is also a key signature of a multifractal intermittent turbulence phenomenology. It suggests that   dissipation, in the sense of conversion of magnetic to kinetic energy, in collisionless reconnection on kinetic scales has an analogue in dissipation in fluid-like turbulent phenomenology, in that it proceeds via a spatial multifractal field of structures generated by an intermittent cascade. If this is the case, then the spatial dissipation field will also be multifractal.

\begin{figure}[!ht]
\begin{center}
\includegraphics[scale=0.2]{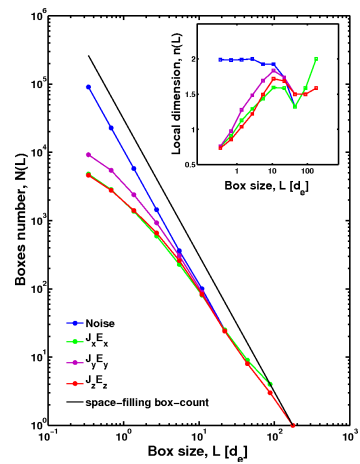}
\caption{ \textbf{Box-counting method.} Plot of the number of boxes, N($L$), versus the box size $L$ within a reconnection dominated turbulent region (red square in Fig.\ref{Bzgrid}) and a noise region (yellow square in Fig.\ref{Bzgrid}) for each component of $\bf J \cdot E$. The inset figure shows the corresponding local dimension, n($L$) = d ln(N($L$))/d ln($L$), against the box size $L$.}
\label{boxcounting}
\end{center}
\end{figure}

We now test this idea by directly quantifying the spatial topology of the field  $\bf J \cdot E$. We show results for the three components of $\bf J \cdot E$,  at  the same time-space slice of the simulation discussed above. We perform the classical box counting method to calculate both global and local fractal dimension \cite{Mandelbrot1977} for a reconnection dominated turbulent region and a region where noise is dominant. The box-counting method consists of dividing a spatial region into boxes of size $L$ using a regular grid, and then counting the number of boxes $N(L)$ that contain non-zero values of a discretized spatial field. We consider the magnitude of each component of $\bf J \cdot E$ and the discretized values are non-zero where it exceeds a threshold, we test the robustness of our results by varying this threshold. For sufficiently small thresholds, this yields the topology of the noise field, but for thresholds above the noise, we obtain the topology of the turbulence. Box counting the thresholded turbulent field then gives its spatial topology in the absence of noise, without the need of a filtering or averaging.  This method can thus probe spatial structures on length scales where the noise power dominates the overall signal power provided that at least in some locations, the signal is above the noise threshold. For fractal geometries N($L$) is expected to depend linearly upon $L$, while non-linear trends of N($L$) against $L$ indicate a multifractal field. In Figure \ref{boxcounting} we plot N($L$) versus $L$ within the reconnection dominated turbulent region and the noise region indicated respectively by the red and yellow squares in Figure \ref{Bzgrid}. The noise region shows a linear behaviour of $N(L)$ with $L$ indicating that the PIC noise is fractal, on the contrary, in the turbulent region the plot introduces curvature for scales L smaller than $\sim 25 d_e$. The inset of Figure \ref{boxcounting} also shows how the local dimension, n($L$) = d ln(N($L$)) / d ln($L$), varies with the scale $L$. Within the noise region, the fractal dimension is roughly constant as $L$ varies, while it changes with the scale $L$ in the turbulent region again, a signature of multifractality. Thus dissipation, in the sense of energy transfer to the plasma via $\bf J \cdot E$, occurs in a spatially intermittent manner.

In conclusion, recent fully kinetic (PIC) simulations in 3D reveal that reconnection is dominated by magnetic structures on multiple scales which manifest highly variable dynamics suggestive of turbulence. We have quantified the ensemble averaged statistical properties of the spatial fields of fluctuations in the magnetic field and in energy transfer to the plasma. 
The magnetic field fluctuations are anisotropic and exhibit the hallmarks of finite range \emph{fluid} turbulence; they have non-Gaussian distributions, exhibit Extended Self-Similarity in their scaling and are spatially multifractal. These signatures are recovered quite robustly across the regions in the simulation domain where reconnection is actively generating fluctuations. This potentially offers a new observational test for reconnection regions using in-situ observations, so that for example recent observations of non-Gaussian fluctuations in a turbulent jet \cite{Huang2012} (see also \cite{Chaston2009,Dai2011}) could be tested for ESS and non-linear ratios of exponents as found here. Furthermore, the fact that we also find that the spatial field of $\bf J \cdot E$ is multifractal suggests that the turbulence converts some of the magnetic energy to plasma kinetic energy in a spatially intermittent manner. 
This suggests that dissipation in the sense of energy transfer to the plasma in collisionless reconnection on kinetic scales has an analogue in fluid-like turbulent phenomenology, in that it proceeds via multifractal structures generated by an intermittent cascade. This provides a starting point for theoretical models of heating in collisionless reconnection; it also suggests that existing analytical and quasi-analytical models of reconnection that do not take into account the development of turbulence in the reconnection layer may need revision.

E.L. and S.C.C. acknowledge the UK EPSRC and STFC. W.D., H.K., and V.R. acknowledge partial support from grants DE-SC0004662 and NASA's Heliophysics Theory Program.
Simulations carried out on Kraken with an allocation of advanced computing resources provided by the National Science Foundation at the National Institute for Computational Sciences.
\bibliographystyle{apsrev}
\bibliography{Bibliography.bib}

\end{document}